\documentclass[aps,preprint]{revtex4}%
\usepackage{amsfonts}
\usepackage{amsmath}
\usepackage{amssymb}
\usepackage{graphicx}%
\providecommand{\U}[1]{\protect\rule{.1in}{.1in}}

\begin{document}
\author{}
\title{Recovering information of tunneling spectrum from Weakly Isolated Horizon}
\author{Ge-Rui Chen}\email{chengerui@emails.bjut.edu.cn}
\author{Yong-Chang Huang}
\affiliation{Institute of Theoretical Physics, Beijing University of
Technology, Beijing, 100124, China}

\begin{abstract}
In this paper we investigate the properties of tunneling spectrum
from weakly isolated horizon(WIH). We find that there are
correlations among Hawking radiations from weakly isolated horizon,
the information can be carried out in terms of correlations between
sequential emissions, and the radiation is an entropy conservation
process.
 We generalize Refs.\cite{bcz1,bcz2,bcz3}' results to a more
general spacetime. Through revisiting the calculation of tunneling
spectrum of weakly isolated horizon, we find that Ref.\cite{bcz2}'s
requirement that radiating particles have the same angular momenta
of unit mass as that of black hole is not needed, and the energy and
angular momenta of emitting particles are very arbitrary which
should be restricted only by keeping the cosmic censorship of black
hole.

\end{abstract}

\maketitle

\section{\textbf{Introduction}}
 In the 1970s, Hawking's astounding discovery that black
holes radiate black body spectrum\cite{sk1,sk2} had greatly
stimulated the development of the theory of black hole, and then
four laws of black hole thermodynamics were
established\cite{jdb,jmb}. Hawking radiation gives us new insights
into gravity physics and also provides some hints of quantum
gravity. From Hawking's famous work, people know that black holes
are not the final state of stars, and, with the emission of Hawking
radiation, they could lose energy, shrink, and eventually evaporate
completely. However, because of the quality of purely thermal
spectrum, it also sets up a disturbing and difficult problem: what
happens to information during black hole evaporation? This scenario
is inconsistent with the unitary principle of quantum
mechanics\cite{swk3,swk4,cgc,sdm}. About the year of 2000, Parikh
and Wilczek, contemplating Hawking's heuristic picture of tunneling
triggered by vacuum fluctuations near the horizon, proposed a
semiclassical method to investigate the emission rate by treating
Hawking radiation as a tunneling process \cite{mkp1,mkp3}. This
method considers the back reaction of the emission particle to the
spacetime, and does not fix the background spacetime. They found
that the barrier of tunneling is created by the outgoing particle
itself, and when energy conservation is considered, a non-thermal
spectrum is given, which supports the underlying unitary theory.

In 2009, Refs.\cite{bcz1,bcz2,bcz3} gave more detail discussions
about Parikh and Wilczek's non-thermal spectrum. They found that
there are correlations among sequential Hawking radiations, the
correlations equal to mutual information, and black hole radiation
is an entropy conservation process, which is consistent with
unitarity of quantum mechanics. Their discussions are based on
stationary black holes, and we study this problem for weak isolated
horizon\cite{aa2,aa5,aa7,bk}--a quasi-local defined black hole, and
prove that for this kind of dynamical black holes, the information
is also not lost, and is encoded into correlations between Hawking
radiations. In our analysis Ref.\cite{bcz2}'s requirement that
radiating particles have the same angular momenta of unit mass as
that of black hole is not needed, and the energies and angular
momenta of emitting particles are very arbitrary which should be
restricted only by keeping the cosmic censorship of black hole.

This paper is organized as follows. In Section 2, we review the
tunneling method to get the non-thermal spectrum of weakly isolated
horizon. In Section 3, we investigate the qualities of this
non-thermal spectrum. In the last Section, we give some discussions
and conclusions.

\section{Review Parikh and Wilczek's tunneling spectrum for weakly isolated horizon}
In this section we review the calculation of tunneling spectrum of
WIH. We have some difference from the original
discussion\cite{xnw1}, and strictly follow Parikh and Wilczek's
calculation\cite{mkp1,mkp3} which does no use explicitly the first
law of black hole thermodynamics.

Ref.\cite{aa7} established the first law of weakly isolated horzion
thermodynamics,
\begin{eqnarray}
\delta E=\frac{1}{8\pi}\kappa\delta A+\Omega \delta
J=\frac{1}{2\pi}\kappa\delta S+\Omega \delta J.
\end{eqnarray}
The expressions of the surface gravity, angular velocity and horizon
energy of weakly isolated horizon are given by
\begin{eqnarray}
\kappa=\frac{R^4-4J^2}{2R^3\sqrt{R^4+4J^2}},
\Omega=\frac{2J}{R\sqrt{R^4+4J^2}},
E=\frac{\sqrt{R^4+4J^2}}{2R},\label{P1}
\end{eqnarray}
where $R$ is the horizon radius and  is defined as
\begin{eqnarray}
R\equiv\sqrt{\frac{A}{4\pi}}.
\end{eqnarray}
$A$ is the area of any cross section of the horizon, so the entropy
can also expressed as
\begin{eqnarray}
S=\frac{A}{4}=\pi R^2.
\end{eqnarray}

In the semiclassical limit, we can apply the WKB formula. The
emission rate $\Gamma$ can be given as
\begin{eqnarray}
\Gamma\sim\exp(-2Im\ I),
\end{eqnarray}
where $I$ is the action of the emitting particle.

The imaginary part of the action for a s-wave outgoing positive
energy particle, from $r_{in}$ to $r_{out}$, can be given as
\begin{eqnarray}
Im\ I=Im\int_{r_{in}}^{r_{out}}p_r
dr=Im\int_{r_{in}}^{r_{out}}\int_{0}^{p_{r}}dp_r dr.
\end{eqnarray}

From Hamilton's equation of the emitting particle,
\begin{eqnarray}
dp_r=\frac{d\varepsilon}{\dot{r}}
\end{eqnarray}
where $\varepsilon$ is the energy of the emitting particles, we can
get
\begin{eqnarray}
Im\ I=Im\int_{r_{in}}^{r_{out}}\int_0^\omega
\frac{d\varepsilon}{\dot{r}}dr.
\end{eqnarray}
From Ref.\cite{xnw1}, the outgoing geodesic is
\begin{eqnarray}
\dot{r}=B_t(\varepsilon+\overline{\varepsilon})r+O(r^2)=\kappa
r+O(r^2),
\end{eqnarray}
where $\kappa=B_t(\varepsilon+\overline{\varepsilon})$ is the
surface gravity of the horizon, and is constant on the horizon. So
the imaginary part of action is
\begin{eqnarray}
Im\ I&=&Im\int_{r_{in}}^{r_{out}}\int_0^\omega
\frac{d\varepsilon}{\kappa
r+O(r^2)}dr=Im\int_0^\omega\int_{r_{in}}^{r_{out}} \frac{dr}{\kappa
r+O(r^2)}d\varepsilon\nonumber\\
&=&Im\int_0^\omega[\pi
i\frac{1}{\kappa}]d\varepsilon=\pi\int_0^\omega\frac{d\varepsilon}{\kappa},
\end{eqnarray}
where the integral of $r$ is done by deforming the contour around
the pole in the third equality.

For non-rotating WIH, $\Omega=0, J=0$, so we can get from (\ref{P1})
\begin{eqnarray}
\kappa=\frac{1}{4E}.
\end{eqnarray}

We fix the total mass of the space-time, and allow the black hole
mass to fluctuate. After emitting a particle with energy
$\varepsilon$  the black hole mass becomes $E-\varepsilon$, so we
obtain
\begin{eqnarray}
Im\
I=\pi\int_0^\omega\frac{d\varepsilon}{\kappa^{'}}=\pi\int_0^\omega4E^{'}d\varepsilon=\pi\int_0^\omega4(E-\varepsilon)d\varepsilon=4\pi\varepsilon(E-\frac{\varepsilon}{2}).
\end{eqnarray}
According to the definition of entropy of WIH,
\begin{eqnarray}
S=\pi R^2=4\pi E^2,
\end{eqnarray}
we have the change of the entropy after the particle radiates,
\begin{eqnarray}
\Delta
S=4\pi[(E-\varepsilon)^2-E^2]=-8\pi\varepsilon(E-\frac{\varepsilon}{2}).
\end{eqnarray}
So we get the tunneling rate
\begin{eqnarray}
\Gamma=\exp{(-2Im\
I)}=\exp[-8\pi\varepsilon(E-\frac{\varepsilon}{2})]=\exp(\Delta
S)=\exp(4\pi[(E-\varepsilon)^2-E^2]).
\end{eqnarray}

Next, we discuss the rotating WIH. From Eqs.(\ref{P1}), after some
calculation, we can get
\begin{eqnarray}
\Omega&=&\frac{J}{2E(E^2+\sqrt{E^4-J^2})},
\ \kappa=\frac{\sqrt{E^4-J^2}}{2E(\sqrt{E^4-J^2}+E^2)},\nonumber\\
S&=&\pi R^2=2\pi(E^2+\sqrt{E^4-J^2}).
\end{eqnarray}

For axial symmetric WIH, using the formula\cite{xnw1,jz}, the action
$Im \ I$ should be
\begin{eqnarray}
Im \ I&=&Im\int[p_r dr-p_\phi d\phi]=Im\int[p_r-\frac{p_\phi
\dot{\phi}}{\dot{r}}]dr=
Im\int\int\frac{dH-\dot{\phi}dp_\phi}{\dot{r}}dr\nonumber\\
&=&Im\int\int\frac{d\varepsilon-\Omega dj}{\dot{r}}dr=Im\int\int\frac{dr}{\dot{r}}(d\varepsilon-\Omega dj)\nonumber\\
&=&Im\int\frac{\pi i}{\kappa}(d\varepsilon-\Omega
dj)=\pi\int\frac{d\varepsilon-\Omega dj}{\kappa}\label{P4},
\end{eqnarray}
where we consider the s-wave, the particles radiate along the normal
direction of the horizon, so $\dot{\phi}=\Omega$ according to the
relationship  $t^a\widehat{=}Bl^a-\Omega\phi^a$\cite{aa7,xnw1}. This
is the acquirement for the emitting particles, and emitting
particles do not need to have the original angular momentum of unit
mass of black hole(see Ref.\cite{bcz2}).

When particle's self-gravitation is taken into account we should
replace $E$ and $J$ with $E-\epsilon$, and $J-j$, and substitute
into the expression of $\kappa$ and $\Omega$ in the last
Eq.(\ref{P4}), so we get
\begin{eqnarray}
Im \ I&=&\pi\int\frac{d\varepsilon-\frac{J-j}
{2(E-\varepsilon)[(E-\varepsilon)^2+\sqrt{(E-\varepsilon)^4-(J-j)^2}]}dj}
{\frac{\sqrt{(E-\varepsilon)^4-(J-j)^2}}{2(E-\varepsilon)[\sqrt{(E-\varepsilon)^4-(J-\varepsilon)^2}+(E-\varepsilon)^2]}}\nonumber\\
&=&\pi\int\frac{2(E-\varepsilon)[\sqrt{(E-\varepsilon)^4-(J-\varepsilon)^2}+(E-\varepsilon)^2]}{\sqrt{(E-\varepsilon)^4-(J-\varepsilon)^2}}d\varepsilon\nonumber\\
&&-\frac{J-j}{\sqrt{(E-\varepsilon)^4-(J-\varepsilon)^2}}dj\label{P2}.
\end{eqnarray}

We do not need to do the integration directly. The change of back
hole entropy after emitting a particle is
\begin{eqnarray}
\Delta
S=2\pi[(E-\varepsilon)^2+\sqrt{(E-\varepsilon)^4-(J-j)^2}]-2\pi[E^2+\sqrt{E^4-J^2}].
\end{eqnarray}
We can get
\begin{eqnarray}
\frac{\partial (\Delta S)}{\partial
\varepsilon}&=&-4\pi[\frac{(E-\varepsilon)^3}{\sqrt{(E-\varepsilon)^4-(J-j)^2}}+(E-\varepsilon)],\nonumber\\
\frac{\partial(\Delta S)}{\partial
j}&=&2\pi\frac{J-j}{\sqrt{(E-\varepsilon)^4-(J-j)^2}},
\end{eqnarray}
and substitute the results into Eq(\ref{P2}), then we get
\begin{eqnarray}
ImS&=&-\frac{1}{2}[\int\frac{\partial(\Delta
S)}{\partial\varepsilon}d\varepsilon+\frac{\partial(\Delta
S)}{\partial j}dj]\nonumber\\
&=&-\frac{1}{2}\Delta S.
\end{eqnarray}
So the tunneling rate is
\begin{eqnarray}
\Gamma&=&\exp(-2ImS)=\exp(\Delta S)\nonumber\\
&=&\exp(2\pi[(E-\varepsilon)^2+\sqrt{(E-\varepsilon)^4-(J-j)^2}]-2\pi[E^2+\sqrt{E^4-J^2}]),
\end{eqnarray}
and our next discussion is based on this equation.

\section{Information recovery from tunneling spectrum of weakly isolated horizon}
In this section, we investigate the properties of tunneling spectrum
from weakly isolated horizon following Refs.\cite{bcz1,bcz2,bcz3}.
The probability for the emission of a particle with an energy
$\varepsilon_1$ and an angular momentum $j_1$ is
\begin{eqnarray}
\Gamma(\varepsilon_1,j_1)=\exp(2\pi[(E-\varepsilon_1)^2+\sqrt{(E-\varepsilon_1)^4-(J-j_1)^2}]-2\pi[E^2+\sqrt{E^4-J^2}]).
\end{eqnarray}
And the probability for the emission of a particle with an energy
$\varepsilon_2$ and an angular momentum $j_2$ is
\begin{eqnarray}
\Gamma(\varepsilon_2,j_2)=\exp(2\pi[(E-\varepsilon_2)^2+\sqrt{(E-\varepsilon_2)^4-(J-j_2)^2}]-2\pi[E^2+\sqrt{E^4-J^2}])\label{P3}.
\end{eqnarray}
Please note that $\varepsilon_1, j_1$ and $\varepsilon_2,j_2$
represent two different emitting particles, so the expressions
should have the same form.

Let us consider a process. Firstly a particle with energy and
angular momentum $\varepsilon_1, j_1$ emits, and then a particle
with energy and angular momentum $\varepsilon_2, j_2$ radiates, so
the probability for the emission of second particle is
\begin{eqnarray}
\Gamma(\varepsilon_2,j_2|\varepsilon_1,j_1)=\exp(2\pi[(E-\varepsilon_1-\varepsilon_2)^2+\sqrt{(E-\varepsilon_1-\varepsilon_2)^4-(J-j_1-j_2)^2}\nonumber\\-2\pi[(E-\varepsilon_1)^2+\sqrt{(E-\varepsilon_1)^4-(J-j_1)^2}]).
\end{eqnarray}
 which is the conditional probability and is different from the independent
 probability(\ref{P3}).

The emitting probability for two emissions with energies and angular
momenta $\varepsilon_1,j_1$ and $\varepsilon_2,j_2$ successively,
can be deduced as follows
\begin{eqnarray}
\Gamma(\varepsilon_1,j_1,\varepsilon_2,j_2)&\equiv&\Gamma(\varepsilon_1,j_1)\Gamma(\varepsilon_2,j_2|\varepsilon_1,j_1)\nonumber\\
&=&\exp(2\pi[(E-\varepsilon_1)^2+\sqrt{(E-\varepsilon_1)^4-(J-j_1)^2}]-2\pi[E^2+\sqrt{E^4-J^2}])\nonumber\\
&&\times\exp(2\pi[(E-\varepsilon_1-\varepsilon_2)^2+\sqrt{(E-\varepsilon_1-\varepsilon_2)^4-(J-j_1-j_2)^2}\nonumber\\
        &&-2\pi[(E-\varepsilon_1)^2+\sqrt{(E-\varepsilon_1)^4-(J-j_1)^2}])\nonumber\\
&=&\exp(2\pi[(E-\varepsilon_1-\varepsilon_2)^2+\sqrt{(E-\varepsilon_1-\varepsilon_2)^4-(J-j_1-j_2)^2}\nonumber\\
&&-2\pi[E^2+\sqrt{E^4-J^2}])
\end{eqnarray}
The last equality is nothing but
$\Gamma(\varepsilon_{1}+\varepsilon_{2},j_{1}+j_{2})$, so we get
\begin{eqnarray}
\Gamma(\varepsilon_1,j_1,\varepsilon_2,j_2)&\equiv&\Gamma(\varepsilon_1,j_1)\Gamma(\varepsilon_2,j_2|\varepsilon_1,j_1)=\Gamma(\varepsilon_{1}+\varepsilon_{2},j_{1}+j_{2}).
\end{eqnarray}
This is an important relationship which tells us that the
probability of two particles emitting successively with energies and
angular momenta $(\varepsilon_1,j_1)$ and $(\varepsilon_2,j_2)$ is
the same as the probability of a particle with an energy and angular
momentum $(\varepsilon_1+\varepsilon_2,j_1+j_2)$. And it is easy to
see that
\begin{eqnarray}
\Gamma(\varepsilon_1,j_1,\varepsilon_2,j_2,\cdots,\varepsilon_i,j_i)
&=&\Gamma(\varepsilon_1,j_1)\Gamma(\varepsilon_2,j_2|\varepsilon_1,j_1)\times\cdots\times
\Gamma(\varepsilon_i,j_i|\varepsilon_1,j_1,\cdots,\varepsilon_{i-1},j_{i-1})\nonumber\\
&=&\Gamma(\varepsilon_1+\cdots+\varepsilon_i,j_1+\cdots+j_i),\label{P5}
\end{eqnarray}
which is an important relationship we will use later.

The function
\begin{eqnarray}
C(A\cup B ; A,B)=\ln\Gamma(A\cup B)-\ln[\Gamma(A)\Gamma(B)]
\end{eqnarray}
is used to measure the statistical correlation between two events
$A$ and $B$. For the Hawking radiation, the correlation between the
two sequential emissions\cite{bcz1,bcz2,bcz3,maa} can be calculated
as
\begin{eqnarray}
\ln\Gamma(\varepsilon_1+\varepsilon_2,j_1+j_2)-\ln[\Gamma(\varepsilon_1,j_1)\Gamma(\varepsilon_2,j_2)]
&=&\ln\frac{\Gamma(\varepsilon_1+\varepsilon_2,j_1+j_2)}{\Gamma(\varepsilon_1,j_1)\Gamma(\varepsilon_2,j_2)}\nonumber\\
&=&\ln\frac{\Gamma(\varepsilon_1,j_1)\Gamma(\varepsilon_2,j_2|\varepsilon_1,j_1)}{\Gamma(\varepsilon_1,j_1)\Gamma(\varepsilon_2,j_2)}\nonumber\\
&=&\ln\frac{\Gamma(\varepsilon_2,j_2|\varepsilon_1,j_1)}{\Gamma(\varepsilon_2,j_2)}\neq0,
\end{eqnarray}
which shows that the two emissions are statistically dependent, and
there are correlations between sequential Hawking radiations of WIH.

The conditional probability
$\Gamma(\varepsilon_i,j_i|\varepsilon_1,j_1,\cdots,\varepsilon_{i-1},j_{i-1})$
is the tunneling probability of a particle emitting with energy and
angular momentum ($\varepsilon_i,j_i$) after a sequence of radiation
from $1 \rightarrow (i-1)$, and conditional entropy taken away by
this tunneling particle is then given by
\begin{eqnarray}
S(\varepsilon_i,j_i|\varepsilon_1,j_1,\cdots,\varepsilon_{i-1},j_{i-1})&=&-\ln\Gamma(\varepsilon_i,j_i|\varepsilon_1,j_1,\cdots,\varepsilon_{i-1},j_{i-1}).
\end{eqnarray}

The mutual information for the emission of two particles with
energies and angular momenta $(\varepsilon_1,j_1)$ and
$(\varepsilon_2,j_2)$ is defined as\cite{bcz1,bcz2,bcz3}
\begin{eqnarray}
S(\varepsilon_2,j_2:\varepsilon_1,j_1)&\equiv&
S(\varepsilon_2,j_2)-S(\varepsilon_2,j_2|\varepsilon_1,j_1)\nonumber\\
&=&-\ln\Gamma(\varepsilon_2,j_2)+\ln\Gamma(\varepsilon_2,j_2|\varepsilon_1,j_1)\nonumber\\
&=&\ln\frac{\Gamma(\varepsilon_2,j_2|\varepsilon_1,j_1)}{\Gamma(\varepsilon_2,j_2)},
\end{eqnarray}
which shows that mutual information is equal to correlation between
the sequential emissions, that is to say, information can be carried
out by correlations between Hawking radiations.

Let us calculate the entropy carried out by Hawking radiations. The
entropy of the first emission particle with an energy and angular
momentum $\varepsilon_1,j_1$ is
\begin{eqnarray}
S(\varepsilon_1,j_1)&=&-\ln\Gamma(\varepsilon_1,j_1).
\end{eqnarray}
The conditional entropy of the second emission after the first
emission is
\begin{eqnarray}
S(\varepsilon_2,j_2|\varepsilon_1,j_1)=-\ln\Gamma(\varepsilon_2,j_2|\varepsilon_1,j_1).
\end{eqnarray}
So the total entropy carried by the two emissions becomes
\begin{eqnarray}
S(\varepsilon_1,j_1,\varepsilon_2,j_2)=S(\varepsilon_1,j_1)+S(\varepsilon_2,j_2|\varepsilon_1,j_1).
\end{eqnarray}

Assuming the black hole exhausts after radiating $n$ particles, we
have the relationship
\begin{eqnarray}
\sum_i^n\varepsilon_i=E, \sum_i^nj_i=J,
\end{eqnarray}
where $E,J$ are the mass and angular momentum of the WIH. The
entropy carried out by all the emitting particles is
\begin{eqnarray}
S(\varepsilon_1,j_1,\cdots,\varepsilon_{n},j_{n})&=&\sum_{i=1}^{n}S(\varepsilon_i,j_i|\varepsilon_1,j_1,\cdots,\varepsilon_{i-1},j_{i-1})\nonumber\\
&=&S(\varepsilon_1,j_1)+S(\varepsilon_2,j_2|\varepsilon_1,j_1)\nonumber\\
&&+\cdots+S(\varepsilon_n,j_n|\varepsilon_1,j_1,\cdots,\varepsilon_{n-1},j_{n-1})\nonumber\\
&=&-\ln\Gamma(\varepsilon_1,j_1)-\ln\Gamma(\varepsilon_2,j_2|\varepsilon_1,j_1)-\cdots\nonumber\\
&&-\ln\Gamma(\varepsilon_n,j_n|\varepsilon_1,j_1,\cdots,\varepsilon_{n-1},j_{n-1})\nonumber\\
&=&-\ln[\Gamma(\varepsilon_1,j_1)\times\Gamma(\varepsilon_2,j_2|\varepsilon_1,j_1)
\times\cdots\times\Gamma(\varepsilon_n,j_n|\varepsilon_1,j_1,\cdots,\varepsilon_{n-1},j_{n-1})]\nonumber\\
&=&-\ln\Gamma(\varepsilon_1+\varepsilon_2+\cdots+\varepsilon_n,j_1+j_2+\cdots+j_n)\nonumber\\
&=&-\ln\Gamma(M,J)=2\pi(E^2+\sqrt{E^4-J^2})=S_{WIH},
\end{eqnarray}
where we use the Eq.(\ref{P5}) in the fifth equation. The result
shows that the entropy carried out by all the emitting particles is
equal to the black hole entropy, so the total entropy is conserved.

We have two comments in the above analysis. Firstly, the energy and
angular momentum of emitting particles are not arbitrary, because
the back hole should satisfy the cosmic censorship at any time, that
is to say, the black hole should satisfy $E^4\geq J^2$. If the
extreme case $E^4=J^2$ is reached, the radiation will stop because
the temperature is zero and the sum of the entropy carried out by
Hawking radiation and the remaining entropy of black hole is also
conserved. Secondly, Ref.\cite{bcz2} requires that emitting
particles have the same angular momentum of unit mass as that of
black hole. However we find that the calculation of Parikh and
Wilczek's tunneling spectrum does not need this condition. For the
s-wave, the particles should radiate along the normal direction of
the horizon and the emitting particles' angular velocity should be
equal to the angular velocity of black hole $\Omega$, so there is no
such constrain on emitting particles' angular momenta.
\section{\textbf{Summary and Conclusions}}

\ \ In this paper we generalize the stationary results to weakly
isolated horizon--a dynamical black hole, and find that the
nonthermal spectrum of weakly isolated horizon also has correlation,
information can be carried out by such correlations, and the entropy
is conserved during the radiation process. In our analysis we find
that the emitting particle's angular momentum is very general and is
restricted only by keeping the cosmic censorship of black hole.

\begin{acknowledgements}
This work is supported by National Natural Science Foundation of
China (No. 11275017 and No. 11173028)
\end{acknowledgements}

\end{document}